\documentclass[conference]{IEEEtran}
\usepackage{cite}
\usepackage{amsmath,amsfonts,amssymb,amsthm }
\usepackage{mathtools,xparse}

\usepackage{url}
\usepackage[]{fancyhdr} %

\ifCLASSOPTIONcompsoc
 \usepackage[caption=false,font=normalsize,labelfont=sf,textfont=sf]{subfig}
\else
 \usepackage[caption=false,font=footnotesize]{subfig}
\fi

\usepackage{multirow}
\usepackage{tabularx}
\usepackage{booktabs}
\usepackage{threeparttable}
\usepackage{siunitx}
\usepackage[table,x11names]{xcolor}
\usepackage{bbm}
\usepackage{balance}
\usepackage{standalone}
\usepackage{nicematrix}

\usepackage{nomencl}
\usepackage{algorithm}
\usepackage{algpseudocode}

\usepackage{graphicx}
\usepackage{array}
\usepackage[table]{xcolor}

\usepackage{hyperref}
\usepackage[capitalise]{cleveref}

\crefformat{equation}{(#2#1#3)}

\usepackage{DTUColorScheme}
\usepackage[edges]{forest}
\usepackage{myPlotStyle}

\newcommand*{\myDots}{\ifmmode\mathellipsis\else.\kern-0.13em.\kern-0.13em.\fi} 

\IEEEoverridecommandlockouts

\begin{document}



\title{A Dataset Generation Toolbox for Dynamic Security Assessment: On the Role of the Security Boundary\\
\thanks{The work was supported by the ERC Starting Grant VeriPhIED, funded by the European Research Council, Grant Agreement 949899.}}

\author{\IEEEauthorblockN{Bastien Giraud\IEEEauthorrefmark{1}, Lola Charles, Agnes Marjorie Nakiganda, Johanna Vorwerk, and Spyros Chatzivasileiadis}
\IEEEauthorblockA{Department of Wind and Energy Systems, Technical University of Denmark, Lyngby, Denmark}
\IEEEauthorblockA{\IEEEauthorrefmark{1}bagir@dtu.dk}
}

\maketitle
\thispagestyle{fancy}
\pagestyle{fancy}

\begin{abstract}
Dynamic security assessment (DSA) is crucial for ensuring the reliable operation of power systems. However, conventional DSA approaches are becoming intractable for future power systems, driving interest in more computationally efficient data-driven methods. Efficient dataset generation is a cornerstone of these methods. While importance and generic sampling techniques often focus on operating points near the system's security boundary, systematic methods for sampling in this region remain scarce. Furthermore, the impact of sampling near the security boundary on the performance of data-driven DSA methods has yet to be established. This paper highlights the critical role of accurately capturing security boundaries for effective security assessment. As such, we propose a novel method for generating a high number of samples close to the security boundary, considering both AC feasibility and small-signal stability. Case studies on the PGLib-OPF 39-bus and PGLib-OPF 162-bus systems demonstrate the importance of including boundary-adjacent operating points in training datasets while maintaining a balanced distribution of secure and insecure points. 


\end{abstract}

\begin{IEEEkeywords}
Security boundary, data generation, dynamic security assessment, machine learning, power system operation
\end{IEEEkeywords}

\IEEEpeerreviewmaketitle

\section{Introduction}
The green energy transition and the broad deployment of converter-interfaced resources introduce faster and more complex dynamics in power systems worldwide that challenge power system reliability, security, and stability. To address the growing uncertainty in power generation due to the integration of intermittent renewable energy sources \cite{panciatici2012operating}, one potential solution is to evaluate all possible operating points (OPs) and severe contingencies offline. However, assessing these OPs against multiple contingencies through time-consuming simulations and high-fidelity system modeling \cite{konstantelos2016implementation} becomes intractable as future power systems incorporate more distributed generation and variable prosumer behavior \cite{zafar2018prosumer}.

As an alternative to conventional methods, data-driven approaches for dynamic security assessment (DSA) are gaining traction \cite{duchesne2020recent, marot2022learning}.
DSA generally measures the ability of a power system to withstand a pre-defined set of contingencies by reaching a stable post-fault state.
Data-driven techniques aim to address the challenges of conventional DSA by leveraging offline simulations and machine learning (ML) to create efficient proxies for both offline and online decision-making \cite{duchesne2020recent}. For example, the authors in \cite{krishnan2011efficient} train a decision tree (DT) to derive operating rules for voltage collapse issues on the French transmission grid, classifying OPs based on whether they meet the required voltage margin. Similarly, \cite{bugaje2023generating} trains a DT for transient stability analysis, differentiating between OPs where the rotor angle difference remains below $180^\circ$ (stable) or exceeds $180^\circ$ (unstable) 10 seconds after fault clearance. In general, these methods classify OPs as \emph{secure} or \emph{insecure}, depending on the specific stability criterion under consideration. However, the performance of such ML tools heavily relies on the quality of the training dataset \cite{bugaje2023generating}. 

One way to acquire datasets is by utilizing historical data. The proliferation of phasor measurement units and advanced monitoring tools in control centers has significantly increased data availability \cite{sevilla2022state}. However, since power systems predominantly operate in stable conditions, most recorded OPs reflect stable states. As a result, such datasets are often biased and lack representation of rare but critical OPs, creating gaps in classification accuracy for a security assessment \cite{duchesne2020recent}. To overcome this limitation, synthetic data generation is vital in constructing balanced datasets that include both secure and insecure OPs.



State-of-the-art methods for generating synthetic datasets for DSA are broadly categorized into three approaches: (1) historical sampling, (2) importance sampling, and (3) generic sampling. Historical sampling methods, such as the vine-copula-based approach in \cite{konstantelos2018using} and the composite modeling approach in \cite{sun2016evaluating}, leverage historical data to account for the dependent structure of variables. However, they are limited in their ability to generalize to OPs outside the historical records. In contrast, importance sampling methods \cite{krishnan2011efficient, liu2013systematic, liu2013importance} aim to enhance the representation of rare events by introducing a bias in the sampling process, thereby concentrating more OPs near the security boundary. However, the input space is typically unbalanced, since the insecure region is significantly larger than the secure region. Thus, despite the advantages that importance sampling entails, obtaining a sufficient number of secure samples close to the boundary remains challenging as highlighted in \cite{thams2019efficient}. To obtain more secure samples, generic sampling methods, such as those in \cite{bugaje2023split, joswig2022opf}, aim to maximize coverage of the operating space defined by AC power flow equations. But the computational complexity of sampling in high-dimensional spaces poses significant challenges even for high performance computation methods. Although \cite{krishnan2011efficient, liu2013systematic, liu2013importance, bugaje2023generating} emphasize the importance of sampling near the security boundary to consider information-rich areas, there is still no scientific evidence in the literature confirming that these boundary-adjacent samples actually enhance data-driven DSA methods.

In this work, we address the identified research gaps by identifying the critical role of sampling near the security boundary to enhance the accuracy and efficiency of ML classifiers for DSA. Specifically, we demonstrate that an ML agent trained with a higher proportion of samples near the security boundary can still accurately classify OPs far from this boundary. To obtain OPs close to the security boundary, we extend the sampling approach in \cite{thams2019efficient, venzke2021efficient, charles2024finalcode}. More specifically, the presented sampling method tackles the challenge of sampling in high-dimensional spaces by employing optimization-based bound tightening (OBBT) techniques from \cite{boundtight} to systematically reduce the feasible operating space. Further significant reduction is achieved through convex relaxations and separating hyperplanes, as discussed in \cite{venzke2021efficient}. Each OP is evaluated for small-signal stability, and the directed walk algorithm from \cite{thams2019efficient} is used to identify samples closer to the security boundary. The resulting dataset is constructed to contain samples within a user-specified margin around the security boundary, balanced between secure and insecure OPs, and validated for both AC feasibility and small-signal stability. We then compare our method against naive and importance sampling benchmarks, evaluating performance on a decision tree (DT). Finally, we analyze the misclassified points and assess the need for a balanced dataset. In short, the contributions of the presented work are as follows:
\begin{itemize}
    \item Building up on our previous work, we make for the first time publicly available an efficient dataset generation toolbox (see github link in \cite{giraud_dsa_learn}), combining the separating hyperplane \cite{venzke2021efficient} and directed walk \cite{thams2019efficient} algorithms. We show that our proposed method effectively finds many samples around the security boundary. Our goal with this toolbox is to enable researchers and industry to generate high quality datasets that can accelerate the development and testing of new data-driven methods. It is built in a modular way with the hope that fellow researchers can further develop it and extend it.  
    \item We illustrate that a DT trained on samples around the security boundary can also accurately classify points lying far from the boundary.
    \item We demonstrate the importance of accurately capturing the security boundary by analyzing the misclassified samples, and by comprehensively investigating the distribution of samples of our proposed method and two benchmark approaches.
\end{itemize}
The remainder of this paper is structured as follows. \cref{sec:SA} provides an introduction to DSA and establishes crucial definitions. Then, \cref{sec:Method} details our proposed method for finding and sampling in the vicinity of the security boundary. \cref{sec:Case_studies} analyses the obtained datasets and benchmarks the performance of a DT trained on the dataset obtained from the proposed method against generic and importance sampling methods for two representative test systems. Finally, \cref{sec:Conclusion} concludes the work.

\section{Dynamic Security Assessment} \label{sec:SA}

This Section provides an introduction to DSA of power systems. A secure system is one that respects operational limits under pre-contingency, during-contingency, and post-contingency conditions. DSA assesses if the system remains stable after a contingency and transitions to a stable post-contingency steady-state \cite{cigre2007review}. As such, all aspects of security and stability, including thermal limits, voltage and frequency constraints, and various types of system stability, such as voltage stability, transient stability, and small-signal stability, must be considered. The computations required to assess all these aspects of security for a single operating condition are thus computationally highly demanding. In this work, we focus on one specific stability aspect: small-signal stability assessment. Additionally, we consider the steady-state feasibility of the system, meaning that a stable OP exists for which a power flow solution satisfies the system’s operational constraints. Hence, for the remainder of this paper, the term DSA refers to the assessment of steady-state feasibility and small-signal stability of an OP. 

To clearly differentiate the terms used throughout the paper, we classify an OP as follows:
\begin{itemize}
    \item an OP is \emph{feasible} when it does not violate any operational constraints in steady-state, e.g. all voltage magnitudes remain within bounds and thermal limits are met,
    \item an OP is considered \emph{stable} when it returns back to a stable state following a small-signal disturbance, and finally
    \item an OP is considered \emph{secure} when it is both steady-state feasible and small-signal stable.
\end{itemize}

The remainder of this Section outlines how DSA methods assess feasibility and small-signal stability, before formally defining the stability boundary. 

\subsection{Feasibility}

For an OP to be feasible, a power flow solution must exist that fulfills all physical system constraints, i.e. line flow and voltage magnitude limitations. Finding every single feasible OP for a system is still an open research question. A systematic method, however, that can help us determine feasible OPs, and which we will use in this paper, is through optimization, and specifically the AC Optimal Power Flow (AC-OPF). The AC-OPF tries to determine the most economical operating point that meets all physical system constraints. Hence, all points that satisfy the AC-OPF constraints, i.e. the AC-OPF feasible space, correspond to the entire collection of feasible OPs that we are looking for. 

AC-OPF is, however, a non-linear and non-convex optimization problem. To enhance computational efficiency and generate numerous feasible samples, our methodology incorporates a convex relaxation of the AC-OPF, specifically employing the quadratic convex~(QC) relaxation. We choose the QC relaxation over other alternatives, such as Semi-Definite Programming and Second-Order Cone relaxations, as it provides a good trade-off between computational complexity and the tightness of the relaxation \cite{coffrin2015qc}. However, other convex relaxations could also be applied. A detailed description of the QC and other convex relaxations is provided in \cite{molzahn2019survey}. Here, we provide a brief overview of the applied AC-OPF and QC relaxation. 

\subsubsection{AC Optimal Power Flow}

To efficiently generate a dataset of feasible OPs, the AC-OPF is employed to adjust infeasible OPs and identify the closest feasible solutions. Generally, an AC-OPF minimizes a user-defined objective, typically the fuel cost in a power system context, while ensuring compliance with operational constraints. Let us consider a power system with a set of $\mathcal{N}$ buses. While the subset $\mathcal{G}$ contains generator buses, the subset $\mathcal{D}$ contains buses serving a load. The subset $(i,j) \in \mathcal{L}$ contains power lines that connect bus $i$ to bus $j$. The goal of the optimization is to find the complex bus voltage magnitudes $V_k$ for each bus $k \in \mathcal{N}$ and the complex power dispatch $S_G = P_g +j Q_g$ for every generator $k \in \mathcal{G}$ that satisfy the following operational constraints:
\begin{subequations} \label{eq:acopf_constraints}
    \begin{align}
    \label{eq:voltage_limits}
        (V_k^{\text{min}})^2 \leq V_k(V_k)^\star \leq (V_k^{\text{max}})^2 \quad &\forall k \in \mathcal{N}\\
        \label{eq:generator_limits}
        S_{G_k}^{\text{min}} \leq S_{G_k} \leq S_{G_k}^{\text{max}} \quad&\forall k \in \mathcal{G}\\
        \label{eq:line_limits}
        |S_{ij}| \leq S_{ij}^{\text{max}} \quad &\forall (i, j) \in \mathcal{L}\\
        \label{eq:power_balance}
        S_{G_k} - S_{D_k} = \sum_{k,j \in \mathcal{L}} S_{kj} \quad &\forall k \in \mathcal{N}\\
        \label{eq:line_power_flow}         S_{ij} = (Y_{ij})^\star V_i  (V_i)^\star - (Y_{ij})^\star V_i  (V_j)^\star \quad &\forall (i, j) \in \mathcal{L}\\
        \label{eq:angle_difference} \theta_{ij}^{min} \leq (V_{i}(V_{j})^\star) \leq \theta_{ij}^{max} \quad &\forall (i, j) \in \mathcal{L}
    \end{align}
\end{subequations}
where \eqref{eq:voltage_limits} ensures that the bus voltage magnitudes are constrained by their upper and lower limits, \eqref{eq:generator_limits} bounds the generator's complex power outputs, ~\eqref{eq:line_limits} limits the line flows across each transmission line and \eqref{eq:line_power_flow} formulates the line flow. 
Furthermore, \eqref{eq:power_balance} enforces that generation and demand are balanced at every node and \eqref{eq:angle_difference} constrains the angle difference between the two line ends for each transmission line.

\subsubsection{Quadratic Convex (QC) Relaxation}

The AC-OPF constraints described in \eqref{eq:acopf_constraints} include non-convex terms arising from the product of the voltages (i.e. $V_k(V_k)^\star$), which complicates the optimization problem and motivates the use of convex relaxations to reduce computational complexity. QC relaxations, as proposed in \cite{coffrin2015qc}, employ convex envelopes in the polar form of the AC-OPF to relax the dependencies among voltage variables. Additionally, an auxiliary matrix variable $W$ is introduced to represent the product of the complex bus voltages, where:
\begin{equation} \label{eq:non_convexity}
    W_{ij} = V_i (V_j)^\star.
\end{equation}
This allows for reformulation of \cref{eq:voltage_limits}, \cref{eq:line_power_flow} and \cref{eq:angle_difference} as follows:
\begin{subequations} \label{eq:qc_constraints}
\begin{align}
    \label{eq:voltage_limits_qc}
        (V_k^{\text{min}})^2 \leq (W_{kk}) \leq (V_k^{\text{max}})^2 \quad &\forall k \in \mathcal{N}\\\label{eq:line_power_flow_qcf}
        S_{ij} = (Y_{ij})^\star W_{ii} - (Y_{ij})^\star W_{ij} \quad &\forall (i, j) \in \mathcal{L} \\\label{eq:line_power_flow_qct}
        S_{ji} = (Y_{ij})^\star W_{jj} - (Y_{ij})^\star (W_{ij})^\star \quad &\forall (i, j) \in \mathcal{L}\\
    \label{eq:angle}
        \operatorname{tan}(\theta_{ij}^{\text{min}}) 
        \leq \frac{\Im(W_{ij})}{\Re(W_{ij})} 
        \leq \operatorname{tan}(\theta_{ij}^{\text{max}}) 
        \quad &\forall (i, j) \in \mathcal{L}
\end{align}
\end{subequations}

The non-convexity is captured by the voltage product in \eqref{eq:non_convexity}. To obtain a convex relaxation, \eqref{eq:non_convexity} is removed from the optimization, and new variables are introduced for the voltages $v_i \angle \theta_i \quad \forall i \in \mathcal{N}$ and squared current flows $l_{ij} \quad \forall (i,j) \in \mathcal{L}$ are added. Additionally, the following convex constraints and envelopes are incorporated:
\begin{subequations} \label{eq:qc_relax}
\begin{align}
    \label{eq:w_relax}
        W_{kk} = \langle v_k^{2} \rangle^{T} \quad &\forall k \in \mathcal{N}\\\label{eq:w_relax_ij_r}
        \Re(W_{kk}) = \langle \langle v_i v_j \rangle^{M} \langle \operatorname{cos}(\theta_i - \theta_j)\rangle^{C} \rangle^{M}  \quad &\forall (i, j) \in \mathcal{L}\\\label{eq:w_relax_ij_i}
        \Im(W_{kk}) = \langle \langle v_i v_j \rangle^{M} \langle \operatorname{sin}(\theta_i - \theta_j)\rangle^{S} \rangle^{M}  \quad &\forall (i, j) \in \mathcal{L}\\\label{eq:current_relax}
        S_{ij} + S_{ji} = Z_{ij} l_{ij} \quad &\forall (i, j) \in \mathcal{L}\\\label{eq:current_relax2}
        |S_{ij}|^2 \leq W_{ii} l_{ij} \quad &\forall (i, j) \in \mathcal{L}
\end{align}
\end{subequations}
The superscripts $T$, $M$, $C$ and $S$ represent convex envelopes for the square, bilinear product, cosine, and sine functions, respectively. $Z_{ij}$ denotes the line impedance. The resulting relaxation of the AC-OPF problem is formulated as a second-order cone program (SOCP) that minimizes a user-defined objective function.

\subsection{Stability}
OPs that are feasible are not necessarily small-signal stable and vice versa. While feasibility ensures that an OP does not violate operational limits, it does not account for the system's ability to remain stable following a disturbance. As such, stability extends beyond feasibility by evaluating the system’s ability to recover and settle at a stable OP after a disturbance. These disturbances can manifest as changes in load, generation, or faults, and may affect different types of stability.
A critical subset that receives more attention due to the increase of inverter-dominated generators is small signal stability. Generally, a disturbance in this context is considered small if the dynamic response of the system is accurately captured when linearizing the differential equations that govern the system response. Typically, in power systems, this includes load and generation fluctuations or generator outages of up to around 0.1~p.u. power with respect to the base power of the entire system.

\begin{figure*}[ht]
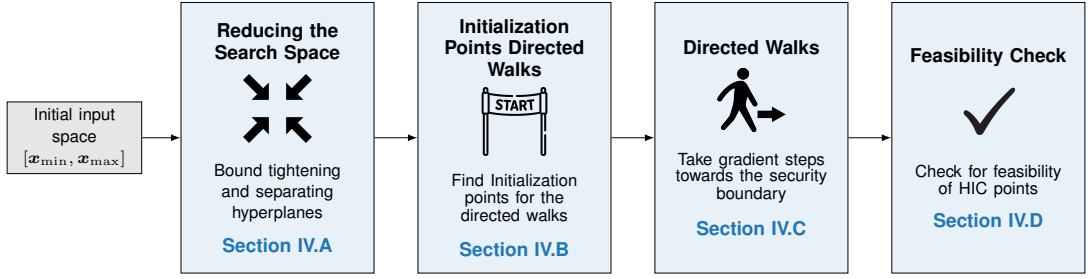

    \centering
    \scalebox{0.9}{\includestandalone{images/methodology/method_workflow}}
    \caption{Flowchart of the proposed methodology.}
    \label{fig:Overview}
\end{figure*}

Small signal stability can be assessed by analyzing the linearized differential equations of a system around a steady-state operating condition, i.e. the state-space representation of the system. The eigenvalues $\lambda_n = \sigma_n + j\omega_n$ of the linearized system state space matrix $\mathbf{A}$, provide insights into system stability. 
Generally, a system is stable if the real part of all eigenvalues of the state-space matrix is negative. Besides the distinction of stability and instability, the real and imaginary parts of the eigenvalue permit an analysis on the oscillation magnitude and duration of an eigenmode after a disturbance: The smaller the real part $\sigma_n$, the faster a post-disturbance oscillation decays. In contrast, the imaginary part $\omega_n$ indicates the oscillation frequency.

The damping ratio~$\zeta$ characterizes the rate at which the magnitude of post-disturbance oscillations diminishes and is calculated as:
\begin{equation}
    \zeta = \frac{-\sigma_n}{\sqrt{\sigma_{n}^{2} + \omega_{n}^{2}}}.
\end{equation}

Consequently, the damping ratio provides valuable insights into the post-disturbance dynamic behavior of that mode: 
\begin{itemize}
    \item Positive damping $\zeta > 0$ indicates a decaying oscillation with rate $\zeta$. As such, the smaller the damping, the slower the oscillation decays.
    \item Zero damping, $\zeta = 0$, indicates a stable, sinusoidal oscillation of the mode in the post-disturbance dynamics.
    \item Negative damping , $\zeta < 0$, indicates an increasing oscillation magnitude over time, rendering the mode, and thus the entire system, unstable. 
\end{itemize}
Note that one unstable mode can cause the system to collapse, and the system is considered small-signal unstable if a single mode is subject to negative damping. Thus, small-signal stability assessment typically aims to identify the least damped mode. In addition, power systems are typically operated with a stability margin, e.g. the damping of the mode closest to the instability boundary should be higher than a pre-defined margin.

The integration of renewable energy sources, such as wind and solar, further challenges small-signal stability due to their intermittency and lower predictability, increasing generation variability. Additionally, renewable generators do not contribute to system inertia, thus lowering overall inertia levels. As such, power systems with high integration of renewable generation are more susceptible to small disturbances because they inherit less inertia to dampen frequency changes. Consequently, small-signal stability analysis is becoming more crucial for early detection of stability issues, particularly in low-carbon grids, to ensure secure, reliable grid operation.

For our proposed sampling technique, a stability index that measures the distance to the stability boundary is required. For example, if the focus was on voltage stability, the voltage margin could be used as a measure. However, determining the voltage margin at every bus requires extensive analysis, thus it is not considered here and should be considered in future work. In contrast, the damping ratio is a computationally efficient, system-wide measure for small-signal stability, motivating its use as the assessment criterion.


\subsection{Security Boundary} \label{sec:SB}

The security boundary $\gamma$ divides the secure from the insecure region. It can correspond to a specific stability boundary, e.g. small-signal or voltage stability. It can also represent a certain stability margin, i.e. OPs not satisfying the margin belonging to the insecure region. It can also be a combination of security indices, e.g. the intersection of AC feasibility and small-signal stability, as it is defined in this work.

The OPs around the stability boundary are potentially extremely valuable for data-driven DSA. As such, the high information content (HIC) region is defined as the set of OPs within a margin $\beta$ of the security boundary. Formally, the HIC region includes all points satisfying:
\begin{equation}
    \Omega = \{ \operatorname{OP}_k \in \Psi | \gamma - \beta < \gamma_k < \gamma + \beta \},
\end{equation}
where $\Omega$ is the set of OPs belonging to the HIC region, $\Psi$ is the total set of OPs and $\gamma_k$ is the value of the chosen stability margin for the $k$-th OP. 

For the remainder of this paper, the security boundary $\gamma$ is represented by the combination of feasibility and a pre-defined level of minimum damping ratio, which corresponds to our stability margin. 
%

\section{Dataset Generation Considering the Security Boundary} \label{sec:Method}

This Section details the methodology for efficiently sampling within the HIC region. The goal is to generate a dataset with a high number of samples in the HIC region while ensuring a balanced distribution of secure and insecure samples. \cref{fig:Overview} provides an overview of the proposed method. The first step reduces the input search space through optimization-based bound tightening and a separating hyperplanes algorithm. The next step obtains initialization points for a directed walk algorithm that is employed to find OPs close to the security boundary. Finally, a feasibility check on the HIC-region samples is performed. The remainder of this Section details the mathematical formulation of each step of the methodology.

To simplify the analysis, the methodology presented here only focuses on feasibility and small-signal stability but can easily be extended to incorporate contingencies, generation or load uncertainty, and other types of stability. The relevant degrees of freedom are defined by the control variables described by input vector $\mathbf{x}$ as:
\begin{equation} \label{eq:variables}
    \mathbf{x} = \begin{bmatrix}
    P_{G_i} \\
    |V_j| \\
    S_{D_k}
    \end{bmatrix} \quad \forall i \in \mathcal{G} \setminus \{\text{slack}\}, \quad \forall j \in \mathcal{G}, \quad \forall k \in \mathcal{D},
\end{equation}

where \( P_{G_i} \) represents the active power generated by each generator \( i \in \mathcal{G} \), excluding the generator at the slack bus, \( |V_j| \) denotes the voltage magnitude of each generator \( j \in \mathcal{G} \), and \( S_{D_k} \) corresponds to the complex power demand at each bus \( k \in \mathcal{D} \). All other states in the AC-OPF are determined by solving the non-linear AC power flow equations with the input vector $\mathbf{x}$. The bounds on $\mathbf{x}$ are specified in \eqref{eq:voltage_limits} and \eqref{eq:generator_limits}. It is important to note that, when using the QC relaxation, the variable $|V_j|$ is substituted with $v_j$.

\subsection{Reducing the Search Space} \label{sec:IVA}
The primary challenge in creating a balanced dataset lies in the exponential growth of control variables as system size and complexity, through uncertainties and potential contingencies, increase. This growth renders naive sampling of the entire input space computationally expensive for large systems. Moreover, as demonstrated in \cite{venzke2021efficient}, a significant portion of the input space, $\mathbf{x} \in \left[\mathbf{x}^\mathrm{min}, \mathbf{x}^\mathrm{max}\right]$, is infeasible. To address this, we employ two methods to reduce the initial input space. First, we apply the OBBT algorithm proposed in \cite{boundtight} to achieve a tighter relaxation of the AC-OPF. Second, we use the separating hyperplanes method from \cite{venzke2021efficient} to pre-classify large portions of the input region as insecure.

\subsubsection{Optimization-Based Bound Tightening} 

The OBBT algorithm has two main objectives. First, it directly tightens the bounds on the input vector $\mathbf{x}$, and second, it improves the QC relaxation, which in turn makes the subsequent separating hyperplane algorithm more efficient. The bounds on the voltage magnitudes and angles significantly affect the tightness of the envelopes in \eqref{eq:qc_relax} and thus influence the quality of the QC relaxation. Through iterative bound tightening, we adjust the voltage magnitude and angle bounds to obtain a tighter relaxation, which also reduces the initial input space.

To achieve this, we employ OBBT tightening as described in \cite{boundtight}, which refines the bounds for voltage magnitudes ($V^{max}$, $V^{min}$) at each bus, as well as the angle differences for each transmission line($\theta_{ij}^{min}$, $\theta_{ij}^{max}$). This is achieved through solving convex optimization problems to identify the minimum and maximum values for each optimization variable within the relaxed problem. The process is repeated for a set number of iterations, resulting in tightened bounds on the input vector $\mathbf{x}$, denoted by $\left[ \mathbf{x}^\mathrm{BT,min}, \mathbf{x}^\mathrm{BT,max} \right]$.

\subsubsection{Infeasibility Certificates}

After the bound tightening process, the input space is further reduced by generating infeasibility certificates using separating hyperplanes, as proposed in \cite{venzke2021efficient}. This method eliminates infeasible regions of the input space in a computationally efficient manner, significantly narrowing the search space. Consider an OP $\hat{\mathbf{x}}$ that is infeasible with respect to the non-convex AC-OPF. We solve an optimization problem to determine the closest dispatch, $\mathbf{x}^\star$, which is feasible with respect to the convex QC relaxation. We achieve this by minimizing the distance between the OP $\hat{\mathbf{x}}$ and the variable $\mathbf{x}$, subject to the QC relaxation:
\begin{subequations} \label{eq:hyperplanes}
    \begin{equation}
        \min_{\substack{\mathbf{x}, \mathbf{S_{}}, \mathbf{S_G}, \mathbf{v}, \mathbf{\theta}, \mathbf{l}, \mathbf{W}, \mathit{R}}} R
    \end{equation}
    \begin{equation} \label{eq:qc_constraints_sum}
        \operatorname{s.t.} \cref{eq:generator_limits} \cref{eq:line_limits} \cref{eq:power_balance}, \cref{eq:qc_constraints}, \cref{eq:qc_relax}, \cref{eq:variables}
    \end{equation}
    \begin{equation} \label{eq:radius}
        \sqrt{\sum_{k \forall \mathcal{X}} {(\mathit{x_k} - \mathit{\hat{x}_k}})^2} \leq R
    \end{equation}
\end{subequations}

If the obtained radius \( R \) is greater than zero, the optimal point \( \hat{\mathbf{x}} \) is infeasible with respect to the convex relaxation. In this case, there is no OP \( \mathbf{x} \) that is closer to \( \hat{\mathbf{x}} \) than the point \( \mathbf{x}^\star \). Using this property, we can construct an infeasibility certificate in the form of a hyperplane to classify a large volume as infeasible. If a non-zero radius is obtained by the optimization in \eqref{eq:hyperplanes}, all vectors $\mathbf{x}$ which comply with the following half-space inequality constraint are infeasible with respect to the AC-OPF constraints \eqref{eq:acopf_constraints}:
\begin{equation}
    \overset{\rightarrow}{\mathbf{n}}^T (\mathbf{x} - \mathbf{x}^\star) < 0
\end{equation}

The normal vector of the hyperplane is defined as \( \overset{\rightarrow}{\mathbf{n}} := \mathbf{x}^\star - \hat{\mathbf{x}} \), where \( T \) denotes the transpose operator. Infeasibility with respect to the QC relaxation constraints, given in \cref{eq:qc_constraints_sum}, guarantees infeasibility with respect to the non-convex AC-OPF constraints in \cref{eq:acopf_constraints}. For a detailed explanation and mathematical proof, the reader is referred to \cite{venzke2021efficient}.

\subsubsection{Convex Polytope Construction} \label{sec:IVB}

Using the method of separating hyperplanes, we employ the algorithm proposed in \cite{venzke2021efficient} to classify large portions of the input space as infeasible. The initial input region defined by $\mathbf{x}$, combined with subsequently constructed hyperplanes, defines a convex polytope represented as $\mathbf{A}\mathbf{x} \leq \mathbf{b}$. When a hyperplane is derived using \eqref{eq:hyperplanes}, it can be incorporated into the convex polytope by appending the row $\mathbf{A_k} := \overset{\rightarrow}{\mathbf{n}}^T$ to matrix $\mathbf{A}$ and the entry $b_k := \overset{\rightarrow}{\mathbf{n}}^T \mathbf{x}^\star$ to vector $\mathbf{b}$. "Hit-and-Run" sampling \cite{kroese2013handbook} provides an efficient method for uniformly sampling points within a convex polytope. This approach enables the iterative construction of hyperplanes by restricting sampling to the currently unclassified region.

The full algorithm from \cite{venzke2021efficient} is outlined in \cref{alg:hyperplanes} to maintain readability. \crefrange{alg:l1}{alg:l3} starts with the initial description of the polytope using the tightened input bounds. In \cref{alg:l6}, for a fixed number of iterations $N_1$ we sample from within the polytope. \cref{alg:l8} checks for the feasibility of the OP. If the OP is infeasible $(R > 0)$, a new hyperplane is constructed and added to the polytope in \crefrange{alg:l11}{alg:l12}. The volume of the unclassified input space is tracked by computing the volume $V^{HP}$ of the convex polytope $\mathbf{A}\mathbf{x} \leq \mathbf{b}$. In \crefrange{alg:l14}{alg:l17}, the algorithm incorporates a stopping criterion, halting when the polytope's volume fails to decrease by more than a user-specified percentage $\tau$ over a user-specified number of consecutive iterations $\eta$.

\begin{algorithm}
\caption{Separating hyperplanes algorithm}
\label{alg:hyperplanes}
\begin{algorithmic}[1]
\State Run bound tightening and obtain $\mathbf{x}^\mathrm{BT,\min}$ and $\mathbf{x}^\mathrm{BT,\max}$ \label{alg:l1}
\State Set iteration count: $k \gets 0$ \label{alg:l2}
\State Initialize unclassified region $\mathbf{A}^{(0)}x \leq \mathbf{b}^{(0)}$: \label{alg:l3}
\State \quad $\mathbf{A}^{(0)} := \left[ \mathbf{I}^{|\mathbf{x}| \times |\mathbf{x}|} - \mathbf{I}^{|\mathbf{x}| \times |\mathbf{x}|} \right]^T$ \label{alg:l4}
\State \quad $\mathbf{b}^{(0)} := \left[ (\mathbf{x}^\mathrm{BT,\max})^T \, (\mathbf{x}^\mathrm{BT,\min})^T \right]^T$ \label{alg:l5}
\While{$k \leq N_1$} \label{alg:l6}
    \State Draw random $\mathbf{x}^{(k)}$ from inside $\mathbf{A}^{(k)}\mathbf{x} \leq \mathbf{b}^{(k)}$ \label{alg:l7}
    \State Solve \cref{eq:hyperplanes} with $\hat{\mathbf{x}} := \mathbf{x}^{(k)}$ and obtain $\mathbf{x}^\star$ \label{alg:l8}
    \If{$R > 0$} \label{alg:l9}
        \State Add hyperplane to reduce unclassified region: \label{alg:l10}
        \State \quad $\mathbf{A}^{(k+1)} = \left[ (\mathbf{A}^{(k)})^T \vec{\mathbf{n}} \right]^T$ \label{alg:l11}
        \State \quad $\mathbf{b}^{(k+1)} = \left[ (\mathbf{b}^{(k)})^T \vec{\mathbf{n}}^T \mathbf{x}^\star \right]^T$ \label{alg:l12}
    \EndIf \label{alg:l13}
    \State Compute polytope volume $V^{HP(k)}$ \label{alg:l14}
    \If{$V^{HP(k)} > (1 - \tau)V^{HP(k-1)}$ for $\eta$ consecutive iterations} \label{alg:l15}
        \State \textbf{break} \label{alg:l16}
    \EndIf \label{alg:l17}
    \State $k \gets k + 1$ \label{alg:l18}
\EndWhile \label{alg:l19}
\end{algorithmic}
\end{algorithm}

\subsection{Initialization Points Directed Walks} \label{sec:IVC}


When the separating hyperplanes algorithm is finished, the sampling continues from within the constructed convex polytope. The purpose is to find initialization points from which directed walks towards the security boundary can be performed. To this extent, a number of $N_2$ samples is drawn from the convex polytope $\mathbf{A}^{(N_1)}\mathbf{x} \leq \mathbf{b}^{(N_1)}$. Then, for each sample the AC power flow equations are solved to check for feasibility. If the sample is feasible, it is added to the feasible set, otherwise to the infeasible set. If the sample is infeasible, another AC power flow enforcing the reactive power limits of the generators is conducted. This step is added because preliminary tests have shown that generator reactive power limits are often the only violation. If the sample is feasible, adjusted setpoints are added to the feasible set. If the sample is still infeasible, the non-convex optimization in \eqref{eq:ac_mapping} is solved to map the sample to the closest feasible dispatch of the AC-OPF.

\begin{subequations} \label{eq:ac_mapping}
    \begin{equation}
        \min_{\substack{\mathbf{x}, \mathbf{V}, \mathbf{S}, \mathbf{S_G}, \mathit{R}}} R
    \end{equation}
    \begin{equation}
        \operatorname{s.t.} \cref{eq:acopf_constraints}, \cref{eq:variables}, \cref{eq:radius}
    \end{equation}
\end{subequations}

The dispatch found in \eqref{eq:ac_mapping} is added to the feasible set. This procedure is repeated for all $N_2$ samples. Next, small-signal stability for all the samples in the feasible and infeasible set is assessed by obtaining the damping ratio of the least damped mode.

\subsection{Directed Walks} \label{sec:DWs}

The feasible samples obtained from the convex polytope serve as initialization points for the directed walk algorithm. \cref{sec:DWs_alg} outlines the directed walk algorithm, \cref{sec:DWs_sens} discusses its use of the damping ratio as a sensitivity measure, and \cref{sec:DWs_HIC} describes the approach for obtaining a high number of samples in the HIC region.

\subsubsection{Directed Walk Algorithm} \label{sec:DWs_alg}

The directed walk algorithm, first applied in \cite{thams2019efficient}, is used to find OPs closer to a user-defined stability index. The algorithm employs a variable step size $\alpha$, which depends on the distance of the OP from the security boundary $\gamma$. The distance $d(\operatorname{OP}_k)$ of an OP to the security boundary is defined as:
\begin{equation}
    d(\operatorname{OP}_k) = |\gamma_k - \gamma|,
\end{equation}
where $\gamma_k$ represents the value of the stability index for OP $\operatorname{OP}_k$. The variable step size $\alpha_k$ is defined as:
\begin{equation}
    \alpha_k =
    \begin{cases} 
        \epsilon_1 \cdot P^{max}, & \text{if} d(\operatorname{OP}_k) > d_1, \\
        \epsilon_2 \cdot P^{max}, & \text{if}, d_1 \geq d(\operatorname{OP}_k) > d_2, \\
        \epsilon_3 \cdot P^{max}, & \text{if}, d_2 \geq d(\operatorname{OP}_k) > d_3, \\
        \epsilon_4 \cdot P^{max}, & \text{otherwise}
    \end{cases}
\end{equation}
The step size $\alpha$ is a function of the generator's maximum capacities $P^{max}$ and scalars $\epsilon_{1-4}$, which depend on the distance $d_{1-3}$ to the security boundary. As the OP approaches the boundary, the step size is gradually reduced to prevent overshooting. Both $\epsilon$ and $d$ are user-defined and are adjustable. The direction of the step is determined by the steepest descent of the metric $d(\operatorname{OP}_k)$. The next OP is found using the step size $\alpha_k$ and the gradient $\nabla d(\operatorname{OP}_k)$:
\begin{equation}
    \operatorname{OP}_{k+1} = \operatorname{OP}_k - \alpha_k \cdot \nabla d(\operatorname{OP}_k)
\end{equation}

This method is compatible with any sensitivity measure that can quantify the distance to a selected stability index. In this work, we use the damping ratio to evaluate small-signal stability.

\subsubsection{Sensitivity Measure} \label{sec:DWs_sens}



For small-signal stability, the direction of the directed walk step is determined by the sensitivity of the damping ratio $\zeta$. More specifically, we use the damping ratio of the least dampened mode of the system. The sensitivity of the damping ratio $\zeta$ to parameter $\rho_i$ can be found analytically by taking the partial derivative of the damping with respect to system parameter $\rho_i$. However, because the computation of this partial derivative is computationally demanding, the damping ratio sensitivity of $\zeta$ to $\rho_i$ is determined by a small perturbation of $\rho_i$.

\subsubsection{Sampling the HIC Region} \label{sec:DWs_HIC}

The directed walk algorithm is applied to all $N_2$ samples, with a user-specified number of $\kappa_{\text{max}}$ directed walks performed for each initialization point. When a directed walk enters the HIC region, the following steps are executed:  
\begin{itemize}
    \item[i)] The current OP is stored.  
    \item[ii)] All OPs around this point are analyzed to determine if they also belong to the HIC region. Any additional samples within the HIC region are also stored.  
    \item[iii)] The directed walk continues along one single dimension with the minimal step size until either an OP already in the dataset is encountered, the directed walk exits the HIC region or the directed walk algorithm reaches its limit of $\kappa_{\text{HIC}}$ steps. A discretization interval between generator setpoints of 1MW is used.
\end{itemize}
All unique OPs found within the HIC region are stored for feasibility evaluation. The DW algorithm is highly parallelizable, enabling simultaneous processing of multiple OPs.

\subsection{Final Feasibility Check} \label{sec:IVE}


When the directed walk algorithm is used on all initialization points, i) a feasibility check is performed on all the samples found in the HIC region. Since only generator active power setpoints are modified during the directed walk algorithm, many resulting OPs exhibit minor violations, including voltage magnitude, reactive power limits, or active power limits of the generator assigned to the slack bus. For all infeasible samples in the HIC region which are found by the directed walk algorithm, ii) \eqref{eq:ac_mapping} is solved to find the closest dispatch AC feasible dispatch. Afterwards, the damping ratio of the lowest dampened mode is recomputed. After this final step, the dataset contains many unique OPs located around the security boundary and which is balanced between secure and insecure OPs.

\section{Case Studies} \label{sec:Case_studies}
This Section provides a comprehensive analysis of datasets generated by our proposed method and two benchmark methods—a naive sampling method and an importance sampling method—to illustrate the impact of a high share of OPs in the HIC region. Furthermore, we train DTs for two test systems to assess the performance of a data-driven DSA tool when using a dataset with a high share of HIC OPs, compared to datasets derived from the benchmarks. As such, this Section first outlines the case study settings and the two benchmark sampling methods, before detailing the implementation of the sampling methods. Then, a data analysis on the obtained datasets is performed, and the results for the decision tree trained with the different data sets is shown. 

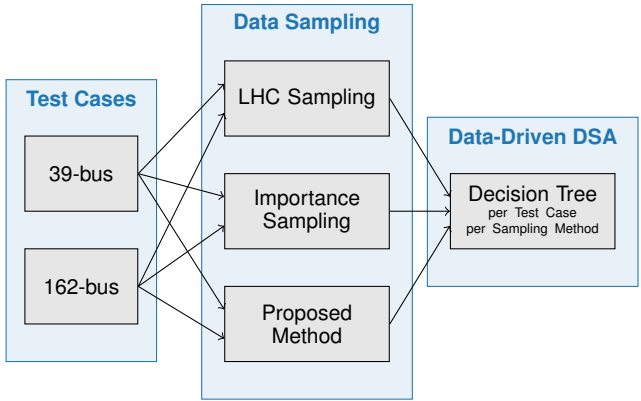
\begin{figure}
    \centering
    \begin{tikzpicture}[scale=1, node distance = 0.5cm]
\footnotesize
\sffamily

\node[block, text width = 1.25cm, align = center](39){39-bus};
\node[block, below of = 39, text width = 1.25cm, node distance = 1.5cm, align = center](162){162-bus};

\node[block, right of = 39, text width = 2cm, xshift = 1cm, yshift = 1cm, align = center] (LHC) {LHC Sampling};
\node[block, below of  = LHC, text width = 2cm, node distance = 1.5cm, align = center] (imp) {Importance Sampling};
\node[block, below of = imp, text width = 2cm, node distance = 1.5cm, align = center](met) {Proposed Method};

\node[block, right of = imp, xshift = 1cm, text width = 2cm, align = center](dt){Decision Tree \\ \tiny per Test Case \\[-0.1cm] per Sampling Method};

\draw[->] (39.east) -- (LHC.170);
\draw[->] (162.east) -- (LHC.190);

\draw[->] (39.east) -- (imp.170);
\draw[->] (162.east) -- (imp.190);

\draw[->] (39.east) -- (met.170);
\draw[->] (162.east) -- (met.190);

\draw[->] (LHC.east) -- (dt.170);
\draw[->] (imp.east) -- (dt.180);
\draw[->] (met.east) -- (dt.190);

\begin{pgfonlayer}{bg}

    \path (39) -- +(-1.,1.25) coordinate (r1);
    \path (162) -- +(1.,-1) coordinate (r2);
    \draw[sBlue ,fill = sBlue!10] (r1) rectangle(r2);
    \node [above of = 39, sBlue, yshift = 0.5cm] { \textbf{Test Cases}};
    
    \path (LHC) -- +(-1.4,1.25) coordinate (r1);
    \path (met) -- +(1.4,-1) coordinate (r2);
    \draw[sBlue ,fill = sBlue!10] (r1) rectangle(r2);
    \node [above of = LHC, sBlue, yshift = 0.5cm] { \textbf{Data Sampling}};
    
     \path (dt) -- +(-1.4,1.25) coordinate (r1);
    \path (dt) -- +(1.4,-1) coordinate (r2);
    \draw[sBlue ,fill = sBlue!10] (r1) rectangle(r2);
    \node [above of = dt, sBlue, yshift = 0.5cm] { \textbf{Data-Driven DSA}};
       
\end{pgfonlayer}

\end{tikzpicture}
    \caption{Description of the datasets generated for dataset evaluation and decision tree training. In total, six datasets are generated and six corresponding decision trees are trained.}
    \label{fig:datasets_case_studies}
\end{figure}


\subsection{Case Study Settings} \label{sec:Case_settings}

We evaluate the proposed method using two test systems from the PGLib-OPF networks (v23.07) \cite{babaeinejadsarookolaee2019power}: the 39-bus and 162-bus systems. The 39-bus system comprises 10 generators and has a peak demand of 6.3~GW, while the 162-bus system includes 12 generators with a peak demand of 7.4~GW. Both systems operate without renewable energy resources, and their nominal load profiles are scaled to $80~\%$ of the nominal load values for this analysis. 

When solving an AC-OPF for both test cases at this loading level, the damping ratio of the most critical mode is slightly above $3\%$. This motivates the definition of the security boundary, which is determined by feasibility and a minimum damping ratio of $\zeta_{min} = 3\%$, a threshold we consider to provide a sufficiently large margin for the damping ratio of the most critical mode. The boundary ensures that the system maintains adequate stability margins under conditions slightly below nominal values (i.e. $80~\%$ loaded). Furthermore, we select a HIC margin value of $\beta = 0.25\%$, resulting in a HIC region defined as $2.75\% < \gamma_{min} < 3.25\%$. This margin is carefully selected during pre-tests to strike a balance between ensuring a sufficiently large HIC region, which allows for effective boundary characterization, and maintaining a narrow scope that facilitates the selective inclusion of OPs. This approach optimizes both the quality and quantity of the resulting dataset. 

All generators in both test systems are modeled as fourth-order round-rotor generators using the RoundRotorQuadratic model from \cite{lara2021powersystems}. Each generator is equipped with an automatic voltage regulator (AVR) of type EXST1, a governor of type TGOV1, and a power system stabilizer (PSS) of type IEEEST, all sourced from the library in \cite{lara2021powersystems}. The specific parameter values used for these models are detailed in \cite{giraud_dsa_learn}. All simulations were performed on 40 cores of the DTU HPC cluster \cite{DTU_DCC_resource}. 

\subsubsection{Benchmark Methods} \label{sec:benchmarks}

To benchmark the proposed method, we employ both a naive sampling method and an importance sampling method. The naive sampling method utilizes Latin Hypercube (LHC) sampling to uniformly sample the entire input space. For each sampled OP, AC power flow calculations are performed to assess feasibility. Infeasible samples are added to the infeasible set, while feasible counterparts are identified by projecting onto the AC feasible region, as described in \eqref{eq:ac_mapping}. Following this, small-signal stability analysis is conducted on both feasible and infeasible samples, thereby classifying the complete dataset.

The importance sampling method, on the other hand, focuses on biasing the sampling process toward the HIC region. Initially, LHC sampling is used to identify OPs within the HIC region. A multivariate normal distribution, $\mathcal{M}(\mu,\Sigma)$, is then fitted to the feasible samples within this region, where $\mu$ represents the mean and $\Sigma$ the covariance matrix of the feasible points. To bias the sampling process towards points closer to the boundary, the covariance matrix $\Sigma$ is scaled by a factor $s < 1$, resulting in a reduced covariance matrix $\Sigma_{red} = s \cdot \Sigma$. In this study, we use $s = 0.25$. Samples are then drawn from the adjusted distribution $\mathcal{M}(\mu,\Sigma_{red})$. For these new samples, AC power flow calculations are performed to determine feasibility. In cases where samples are infeasible, the full AC-OPF is solved to ensure a balanced dataset. Finally, small-signal stability analysis is performed for all the sampled points.

\subsubsection{Dataset Construction Parameters}

For the LHC benchmark, 10~000 samples are uniformly drawn from the entire input space. In contrast, the importance sampling benchmark begins with an initial set of 4~000 samples generated through LHC sampling. This initial set is used to construct a multivariate normal distribution, which then guides the biased sampling process. An additional 10~000 samples are subsequently drawn, with a specific focus on the HIC region. 

For the proposed method, the parameters for the construction of the convex polytope vary for each test system. For the 39-bus system, the polytope is created using either $N_1 = 100$ hyperplanes or until the volume reduction is less than $\tau = 5\%$ for $\eta = 30$ consecutive iterations. For the 162-bus system, the polytope is constructed with up to $N_1 = 400$ hyperplanes or until the same volume reduction of less than $\tau = 5\%$ is observed for $\eta = 100$ consecutive iterations. These parameter values are chosen based on the results found in \cite{venzke2021efficient}. From the resulting polytopes, $N_2 = 4~000$ samples are drawn to initialize the DWs. 

For comparison and analysis, a subset of 10~000 samples is randomly selected from the datasets generated by the proposed method and the two benchmarks. These subsets are then used to perform case studies on two test systems.

\subsection{Implementation} \label{sec:implementation}
The entire framework for the proposed method, as well as for both benchmarks, is developed in Julia. For both AC power flows and AC-OPFs, we rely on Powermodels.jl \cite{coffrin2018powermodels} and JuMP \cite{dunning2017jump}. All optimization problems are solved with the IPOPT solver \cite{wachter2006implementation}. Power system modeling uses PowerSystems.jl \cite{lara2021powersystems}, while small-signal stability analysis employs PowerSimulationsDynamics.jl \cite{lara2023powersimulationsdynamics}. The volume of the convex polytope is computed through the Volesti package \cite{chalkis2020volesti} in R. For the OBBT algorithm we perform three iterations using the OBBT algorithm from \cite{coffrin2018powermodels}. Subsequently, the DW algorithm is executed, with a maximum of $\kappa_{max} = 30$ steps towards the HIC region, followed by up to $\kappa_{HIC} = 15$ steps within the HIC region. The values for $\epsilon_{1-4}$ are empirically tuned and set to $[4,3,2,1]$ respectively. 

We train a total of six DTs to assess the impact of the different sampling techniques on data-driven DSA. For each of the two test systems using datasets from three methods, the proposed method, the LHC sampling benchmark, and the importance sampling benchmark, an individual DT is obtained. Training employs the Classification and Regression Tree (CART) algorithm \cite{breiman1984classification} in Scikit-learn \cite{scikit-learn}, with default settings, Gini impurity as the splitting criterion, and a maximum tree depth of 5. In order to assess the performance of the DTs and to prevent underfitting or overfitting, we apply 10-fold cross-validation while simultaneously monitoring both training and testing accuracies. Additionally, we apply tree pruning with a ccp-alpha score of 0.01 to further improve model generalization. The DTs are employed for binary classification, where a value of 1 indicates a secure OP and a value of 0 indicates an insecure OP. For all six datasets, a 75:25 train-test split is consistently applied for training and evaluating the DTs.

\subsection{Dataset Comparison} \label{sec:analysis}

Before assessing the performance of the decision trees, let us focus on the datasets obtained from the various methods under consideration. As such, this Sections provides quantitative and visual analyses of the datasets constructed by the proposed method and the two benchmark methods. 

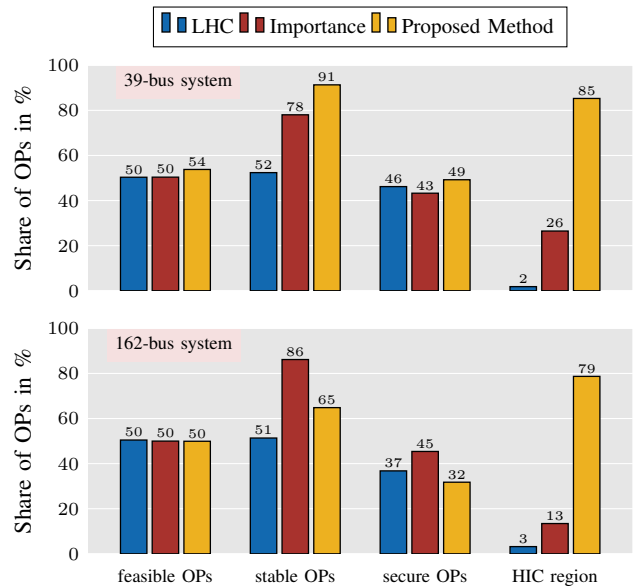
\begin{figure}
    \centering
    \begin{tikzpicture}
\small
\begin{groupplot}[
    group style={
        group name=mygroup, 
        group size=1 by 2,  
        vertical sep=0.5cm,
        horizontal sep=0.2cm,
        xticklabels at=edge bottom,
    },
    legend style={
        at={(0.5,1.17)}, 
        anchor=center,
        legend columns=3
    },
    xlabel={},
    width=7.25cm, 
    height=3cm,
    y tick label style={
        /pgf/number format/.cd,
        fixed,
        precision=5,
        set thousands separator={ },
    },
    every node near coord/.append style={font=\tiny,xshift=0pt,yshift=3pt,anchor=center, /pgf/number format/precision= 0, black},
    nodes near coords align={center},
    nodes near coords,
    xmajorgrids=false,
    ymin = 0, ymax = 100,
    every axis plot/.append style={line width=0.5pt}
]

\pgfplotsset{compat=1.17}

\nextgroupplot[  
    ybar,
    ylabel={Share of OPs in \%},
    symbolic x coords={x1,x2,x3,x4},
    xtick = \empty,
    bar width=10pt,
    enlarge x limits = 0.2,
    y filter/.code={\pgfmathparse{#1/10000*100}} 
]
\addplot[black, fill=pBlue] coordinates {(x1,5034) (x2,5239) (x3,4619) (x4,183)};
\addplot[black, fill=pRed] coordinates {(x1,5038) (x2,7803) (x3,4326) (x4, 2648)};
\addplot[black, fill=pYellow] coordinates {(x1,5379) (x2,9131) (x3, 4923) (x4,8528)};
\legend{LHC, Importance, Proposed Method}

\node[block, fill = sRed!15, anchor = north east, minimum height = 0.1cm, fill opacity = 0.8, text opacity = 1, draw = none, xshift = 1cm] at(axis cs:x1,\pgfkeysvalueof{/pgfplots/ymax}){\scriptsize 39-bus system};

\nextgroupplot[  
    ybar,
    ylabel={Share of OPs in \%},
    symbolic x coords={x1,x2,x3,x4},
    xtick = {x1,x2,x3,x4},
    xticklabels={feasible OPs, stable OPs, secure OPs, HIC region},
    bar width=10pt,
    enlarge x limits = 0.2,
    y filter/.code={\pgfmathparse{#1/10000*100}} 
]
\addplot[black, fill=pBlue] coordinates {(x1,5046) (x2,5135) (x3,3678) (x4,320)};
\addplot[black, fill=pRed] coordinates {(x1,4999) (x2,8616) (x3,4540) (x4, 1343)};
\addplot[black, fill=pYellow] coordinates {(x1,4992) (x2,6483) (x3, 3176) (x4,7869)};

\node[block, fill = sRed!15, anchor = north east, minimum height = 0.1cm, fill opacity = 0.8, text opacity = 1, draw = none, xshift = 1cm] at(axis cs:x1,\pgfkeysvalueof{/pgfplots/ymax}){\scriptsize 162-bus system};

\end{groupplot}
\end{tikzpicture}
    \caption{Share of feasible, stable, secure, and HIC-region samples for the 39-bus system (top) and 162-bus system (bottom). }
    \label{fig:dataset_statistics}
\end{figure}

For the 39-bus system, 110 OPs (2.8\% of the initially sampled points) were identified in the HIC region and used to construct the multivariate normal distribution for importance sampling. Similarly, for the 162-bus system, 138 OPs (3.5\% of the initially sampled points) were identified and utilized for the same purpose. \cref{fig:dataset_statistics} presents the statistics of the 10~000 randomly sampled points from the three datasets for both the 39- and 162-bus systems. 

Since each infeasible sample is paired with a feasible counterpart, all three dataset generation methods ensure a balanced distribution of feasible and infeasible samples, as indicated in the left column of \cref{fig:dataset_statistics}. Both the importance sampling method and the proposed method result in a higher proportion of stable samples relative to unstable ones. For the 39-bus system, this does not affect the balance between secure and insecure samples, as all three methods yield a balanced dataset. However, for the 162-bus system, the LHC sampling and proposed method exhibit a relatively lower proportion of secure samples. The largest difference between the three methods lies when considering the HIC region OPs. For both the 39- and 162-bus test systems, the proposed method achieves a significant representation of samples within the HIC region, capturing $85\%$ and $79\%$ of these samples, respectively.

\begin{figure}[t]
\begin{minipage}[c]{\columnwidth}
    \centering
    \includegraphics[width=0.95\linewidth]{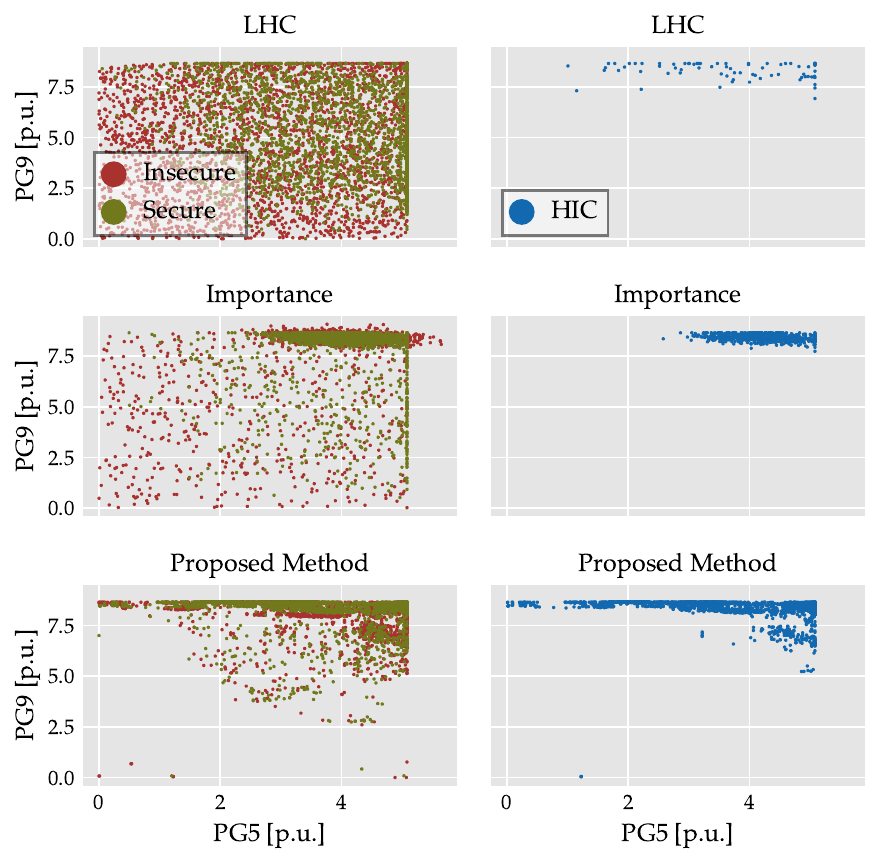}
    \caption{Spread of secure and insecure OPs for the 39-bus system, plotted for generator 5 and 9.}
    \label{fig:39_5_9}
    
    \includegraphics[width=0.95\linewidth]{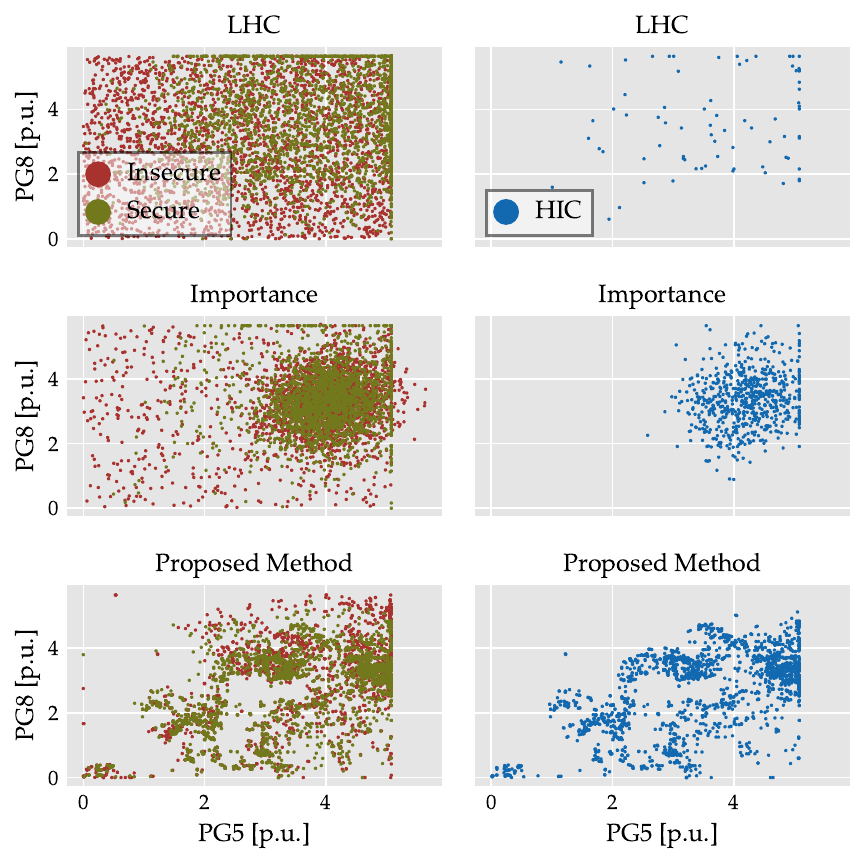}
    \caption{Spread of secure and insecure OPs for the 39-bus system, plotted for generator 5 and 8.}
    \label{fig:39_5_8}
\end{minipage}
\end{figure}

\begin{figure}[t]
\begin{minipage}[c]{\columnwidth}
    \centering
    \includegraphics[width=0.95\linewidth]{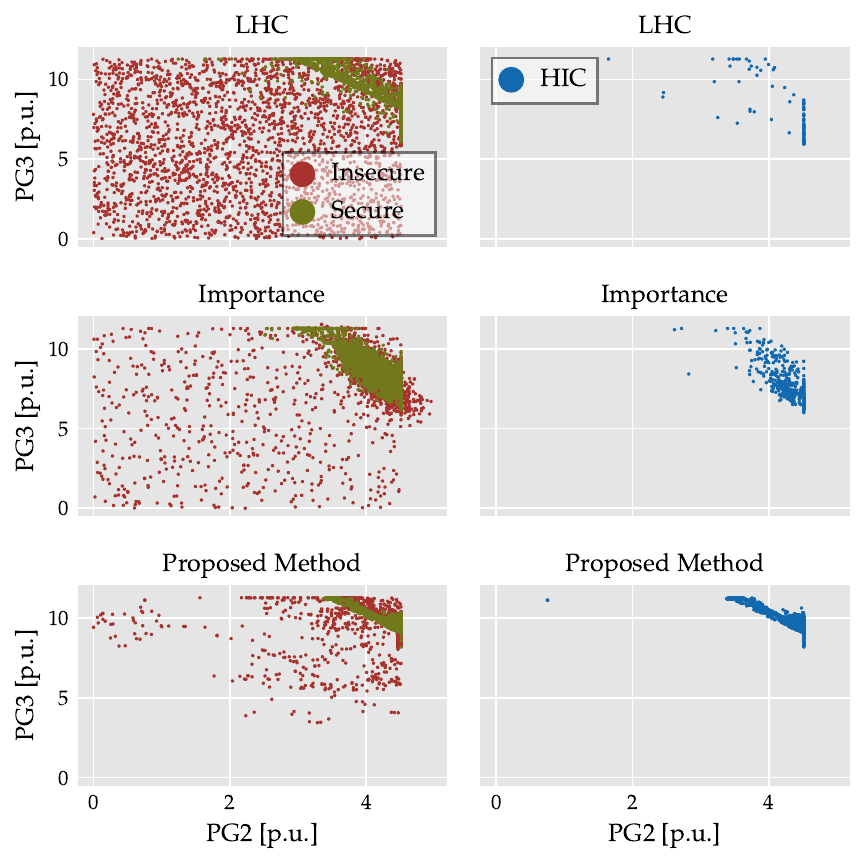}
    \caption{Spread of secure and insecure OPs for the 162-bus system, plotted for generator 2 and 3.}
    \label{fig:162_2_3}
    
    \includegraphics[width=0.95\linewidth]{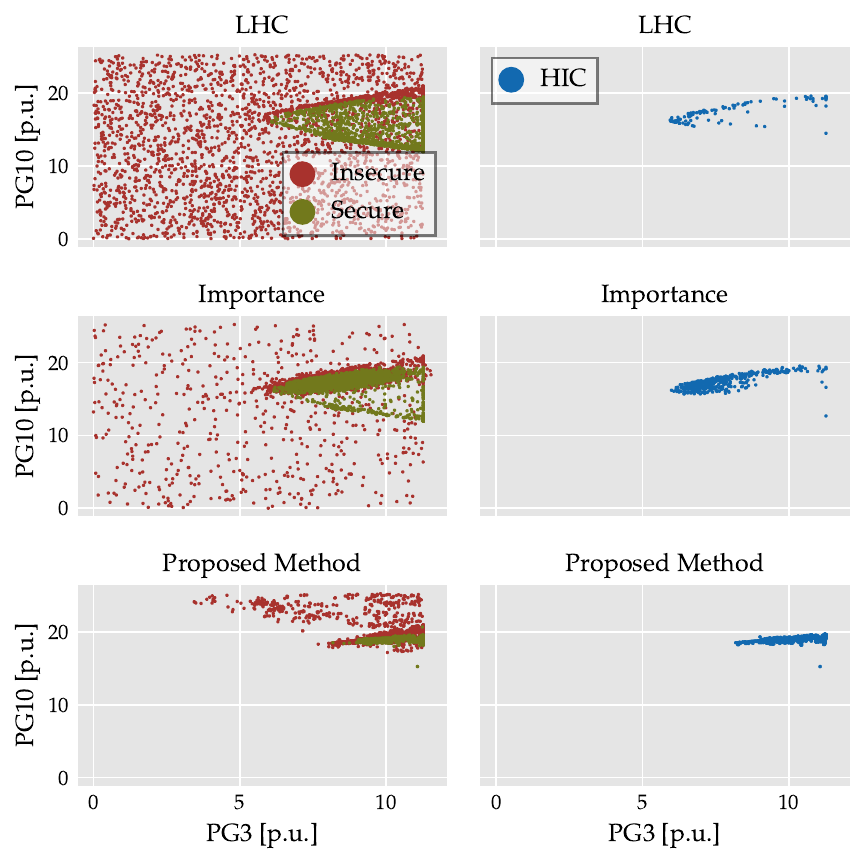}
    \caption{Spread of secure and insecure OPs for the 162-bus system, plotted for generator 3 and 10.}
    \label{fig:162_3_10}
\end{minipage}
\end{figure}

\begin{figure}[t]
    \centering
    \includegraphics[width=0.95\linewidth]{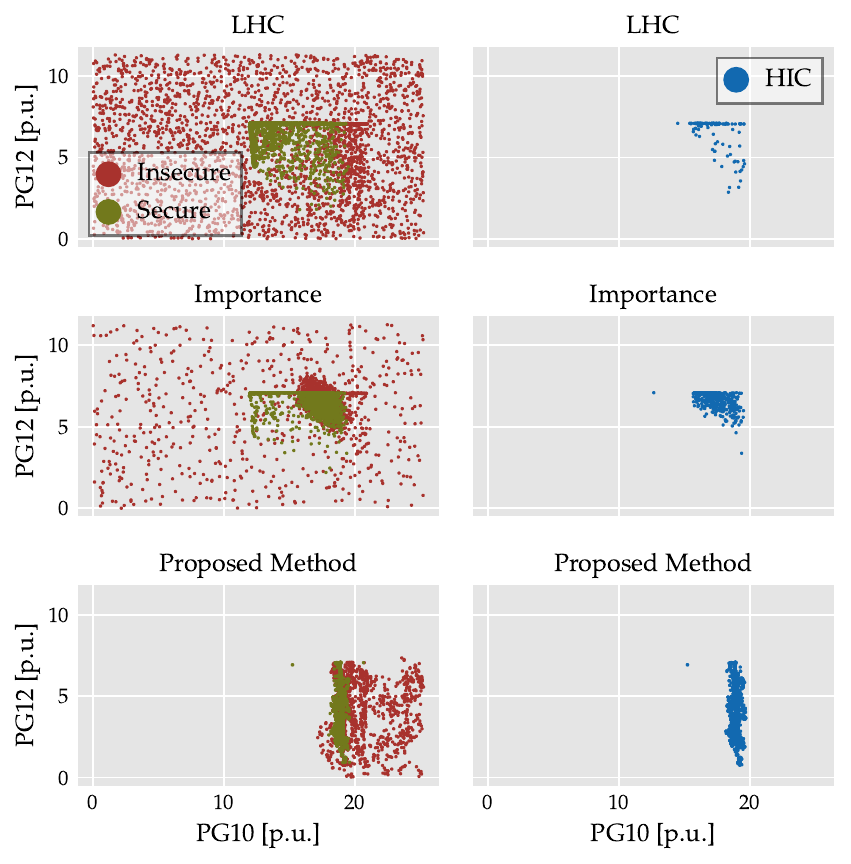}
    \caption{Spread of secure and insecure OPs for the 162-bus system, plotted for generator 10 and 12.}
    \label{fig:162_10_12}
\end{figure}

\crefrange{fig:39_5_9}{fig:162_10_12} provide a visual insight into the distribution of the OPs in two dimensions: the active power setpoints of two generators the respective test cases. Each row in these figures shows the operation points obtained from the mentioned method. The left-hand side depicts the classification into secure and insecure OPs, while the right-hand side illustrates the location of OPs within the HIC region.

\cref{fig:39_5_9} and \cref{fig:39_5_8} present the spread of OPs for the 39-bus system for two different generator sets. Furthermore, \cref{fig:39_5_9} demonstrates that while LHC samples the entire input space, the proposed method and the importance sampling method identify a similar region of importance within the examined dimensions. However, a significant portion of the input space consists of both secure and insecure OPs in close proximity, suggesting considerable overlap between these categories. This overlap is further emphasized in \cref{fig:39_5_8}, which also shows a similar distribution of secure and insecure points across the input space. Moreover, \cref{fig:39_5_8} highlights that the HIC region might be notably dispersed, with no clear boundary emerging that would permit to effectively separate secure from insecure points.

\crefrange{fig:162_2_3}{fig:162_10_12} indicates a more distinct security boundary for the 162-bus system that clearly separates many secure from insecure OPs. The proposed method successfully identified a significant number of OPs within the HIC region, while the LHC method also identified a substantial number of secure and insecure OPs around this region, although not necessarily within the HIC region itself. The importance sampling method similarly identifies many OPs in this area, but its distribution is slightly offset from the boundary, highlighting the challenge of accurately capturing the full security boundary using a naive sampling approach combined with importance sampling. 

Additionally, \cref{fig:162_3_10} presents an example where a distinct secure region is observable, and a significant portion of the input space is insecure. All three methods identified a region of importance at the top of the secure region, with the importance sampling method accurately capturing many points in this area. In contrast, the proposed method captures part of this region but fails to fully capture the entire boundary. 

Finally, \cref{fig:162_10_12} displays an example where a clear boundary between secure and insecure points is visible. The proposed method samples many points at the intersection of secure and insecure areas, whereas the importance sampling method samples points around a specific location but fails to accurately capture the entire boundary. This again demonstrates the difficulty of accurately modeling the security boundary in high-dimensional spaces using a multivariate distribution.

In summary, \crefrange{fig:162_2_3}{fig:162_10_12} demonstrate the effectiveness of the separating hyperplane algorithm in reducing the input search space. Unlike the LHC and importance sampling benchmarks, which sample across the entire input space, the proposed method concentrates on a more constrained region, enabling a more targeted and efficient sampling strategy.

\subsection{Decision Tree Performance} \label{sec:DT}

After analyzing the constructed datasets, this section explores whether a higher number of OPs within the HIC region and a more comprehensive spread of OPs across the security boundary can enhance the performance of DTs for data-driven DSA. As such, the Section first outlines the metric used to evaluate the DTs, before presenting the performance scores for the various DTs. In addition, the location of misclassificatied OPs is anayzed before the results on training with balanced vs. unbalanced datasets are depicted.

\subsubsection{Evaluation Metrics} \label{sec:metrics}
To evaluate the performance of the DTs, the F1-Score is applied:
\begin{equation}
\text{F1-Score} = \frac{2T_p}{2T_p + F_p + F_n}, 
\end{equation}
where, $T_p$, $F_p$, and $F_n$ are true positives, false positives, and false negatives, respectively. The F1-score provides a balanced measure of accuracy by considering both false positives (insecure samples incorrectly classified as secure) and false negatives (secure samples incorrectly classified as insecure). This balance is especially important in DSA, where false positives — misclassifying an insecure operational point as secure — can have severe consequences, potentially causing security threats to the system. 

The F1-score, being the harmonic mean of precision and recall, is particularly valuable when there is a need to balance the correct identification of positive cases (precision) with the ability to identify all actual positive cases (recall). By accounting for both types of misclassification, the F1-score ensures that the model performs well in distinguishing between secure and insecure operational points, providing a more insightful measure of performance than accuracy alone.

\begin{table}[b]
\begin{minipage}[c]{\columnwidth}
    \centering
    \caption{39-bus system F1-scores for the trained DTs.}
    \label{tab:dt_39}
    \renewcommand{\arraystretch}{1.3} 
    \setlength{\tabcolsep}{8pt} 
    \begin{NiceTabular}{w{l}{2cm} cccc}
        \diagbox{\textbf{Training}}{\textbf{Testing}} 
        & LHC & Importance & \Block{}{Proposed\\Method} & Boundary  \\[0.8em] \cline{2-5}
        LHC        & 0.98 & 0.62 & 0.78 & 0.67  \\
        Importance & 0.95 & 0.92 & 0.67 & 0.64  \\
        \rowcolor{lightgray} Method   & 0.95 & 0.91 & 0.92 & 0.77 \\
        \bottomrule
    \end{NiceTabular}
    \bigskip
    
    \centering
    \caption{162-bus system F1-scores for the trained DTs.}
    \label{tab:dt_162}
    \renewcommand{\arraystretch}{1.3} 
    \setlength{\tabcolsep}{8pt} 
    \begin{NiceTabular}{w{l}{2cm} cccc}
        \diagbox{\textbf{Training}}{\textbf{Testing}} 
        & LHC & Importance & \Block{}{Proposed\\Method} & Boundary  \\[0.8em] \cline{2-5}
        LHC   & 0.98 & 0.76 & 0.60 & 0.42 \\
        Importance & 0.97 & 0.97 & 0.64 & 0.48  \\
        \rowcolor{lightgray} Proposed Method   & 0.99 & 0.82 & 0.90 & 0.72 \\
        \bottomrule
    \end{NiceTabular}
\end{minipage}
\end{table}

\subsubsection{DT Performance} \label{sec:performance}

\cref{tab:dt_39} and \cref{tab:dt_162} present the F1-scores of the DTs trained using three different datasets. Each DT is evaluated on the test sets generated by the other two sampling methods, as well as an additional test set named \emph{boundary}. This test set contains OPs generated by the proposed method which lie near the security boundary ($2.9\% < \zeta < 3.1\% $).

For the 39-bus system, the DT trained on the dataset generated by the proposed method generalizes well to the test sets derived from both the LHC and importance sampling methods, i.e. signified by the gray box in \cref{tab:dt_39}. Similarly, the DT trained using the importance sampling dataset generalizes well to the LHC test set. However, the reverse is not observed; the LHC DT does not perform well on the importance test set. Notably, the LHC DT outperforms the importance DT on both the proposed method test set and the boundary test set. Overall, the DT trained on the dataset generated by the proposed method outperformed the other DTs across all test sets on average.

The overall conclusions for the 162-bus system overlap with the previous results. The DT trained on the method test set provides the most stable performance when evaluated across the other datasets. In particular, the DT trained on the proposed method’s dataset achieves strong performance on the LHC test set and outperforms the LHC DT on the importance sampling test set. In contrast, both the LHC and importance DTs show limited generalization capabilities, particularly on the boundary test set. 

\subsubsection{Damping Ratio of Misclassified OPs} \label{sec:misclassification}

\begin{figure}[t]
\begin{minipage}[c]{\columnwidth}
    \centering
    \includegraphics[width=0.95\linewidth]{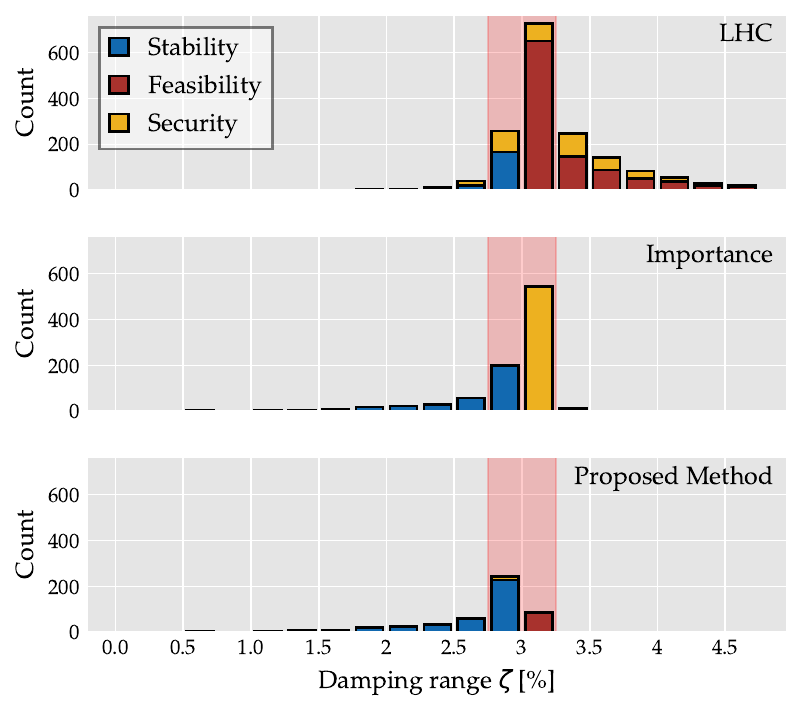}
    \caption{39-bus misclassified OPs and the cause for missclassificaton. The HIC region is highlighted in red.}
    \label{fig:39_missclass}
    
    \centering
    \includegraphics[width=0.95\linewidth]{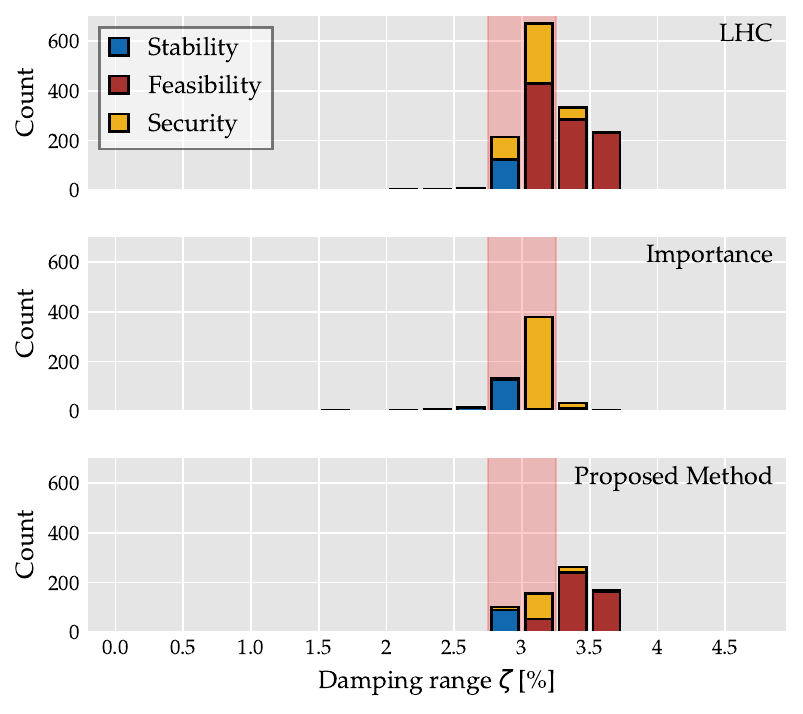}
    \caption{162-bus misclassified OPs and the cause for missclassificaton. The HIC region is highlighted in red.}
    \label{fig:162_missclass}
    \end{minipage}
\end{figure}

\cref{fig:39_missclass} and \cref{fig:162_missclass} provide insights into the misclassification errors of the different DTs for both test systems. For this assessment a large test set, including the points from the LHC, importance sampling, and proposed method sampling, is formed. The diagrams are histograms, where the ordinate depicts the number of misclassified points for a given damping ratio bin. Additionally, the colored section of the bars indicate the cause of misclassification.

Both figures indicate that misclassifications generally occurs near the security boundary, and within the HIC region that is highlighted in red on the figures. The DT trained on the LHC dataset show the highest proportion of misclassified OPs for both the 39- and 162-bus systems. In contrast, the DT trained on the proposed method dataset performs well, accurately classifying many points situated further from the security boundary. This is particularly notable given that $85\%$ and $79\%$ of the samples in the proposed method dataset fall within the HIC region for the 39- and 162-bus systems, respectively.

\cref{fig:39_missclass} and \cref{fig:162_missclass}  further highlight that points with a critical damping ratio exceeding 3~\% are often misclassified due to feasibility or security issues, and rarely solely due to stability. This observation is particularly interesting, as all datasets maintain a nearly balanced distribution between feasible and infeasible points.

\subsubsection{Balancing of Datasets} \label{sec:balance}

\begin{figure}
    \centering
    \begin{tikzpicture}
\small
\begin{groupplot}[
    group style={
        group name=mygroup, 
        group size=1 by 2,  
        vertical sep=0.5cm,
        horizontal sep=0.2cm,
        xticklabels at=edge bottom,
    },
    legend style={
        at={(0.5,1.15)}, 
        anchor=center,
        legend columns=3
    },
    xlabel={},
    width=7.25cm, 
    height=3cm,
    y tick label style={
        /pgf/number format/.cd,
        fixed,
        precision=5,
        set thousands separator={ },
    },
    every node near coord/.append style={font=\tiny,xshift=0pt,yshift=3pt,anchor=center, /pgf/number format/precision= 0, black},
    nodes near coords align={center},
    nodes near coords,
    xmajorgrids=false,
    ymin = 0, ymax = 100,
    every axis plot/.append style={line width=0.5pt}
]

\pgfplotsset{compat=1.17}

\nextgroupplot[  
    ybar,
    ylabel={Share of OPs in \%},
    symbolic x coords={x1,x2,x3,x4},
    xtick = {x1,x2,x3,x4},
    xticklabels={feasible OPs, stable OPs, secure OPs, HIC region},
    bar width=10pt,
    enlarge x limits = 0.2,
    y filter/.code={\pgfmathparse{#1/10000*100}} 
]
\addplot[black, fill=pBlue] coordinates {(x1,5092) (x2,3635) (x3,1647) (x4,7165)};
\addplot[black, fill=pRed] coordinates {(x1,7021) (x2,6210) (x3,5000) (x4, 7994)};

\legend{unbalanced, balanced}

\node[block, fill = sRed!15, anchor = north east, minimum height = 0.1cm, fill opacity = 0.8, text opacity = 1, draw = none, xshift = 1cm] at(axis cs:x1,\pgfkeysvalueof{/pgfplots/ymax}){\scriptsize 162-bus system};

\end{groupplot}
\end{tikzpicture}
    \caption{Share of feasible, stable, secure and HIC region samples for a balanced and unbalanced dataset of the 162-Bus System.}
    \label{fig:statistics162_unbalanced}
\end{figure}
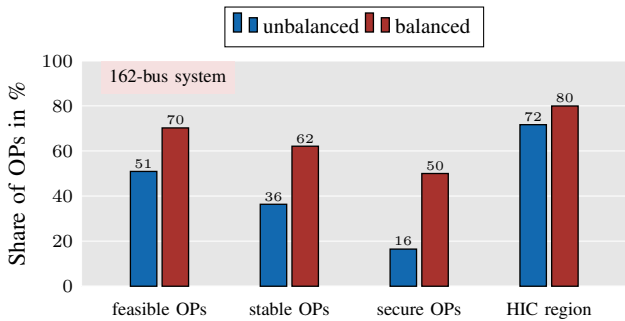

Finally, we study the importance of considering both the security boundary and the balance of the dataset. A new dataset for the 162-bus system, generated by the proposed method, resulted in an unbalanced distribution between secure and insecure samples. One training set was created by randomly sampling 10~000 OPs from this unbalanced dataset. Additionally, a balanced dataset was constructed by equally sampling secure and insecure points from the same dataset. The dataset statistics for both the unbalanced and balanced datasets are presented in \cref{fig:statistics162_unbalanced}. The LHC and importance sampling datasets used for testing are shown in \cref{fig:dataset_statistics}. \cref{tab:dt_162_balanced_unbalanced} displays the results of a DT trained on the balanced and the unbalanced dataset. The results show that not only is the security boundary important for achieving good DT accuracy, but so is the balance between secure and insecure OPs. This is not only true for accuracy on its own test set but also for its generalization capabilities to other test sets.


\begin{table}[t]
    \centering
    \caption{162-bus system F1-scores with an unbalanced and a balanced dataset for the proposed method.}
    \label{tab:dt_162_balanced_unbalanced}
    \renewcommand{\arraystretch}{1.3} 
    \setlength{\tabcolsep}{8pt} 
    \begin{NiceTabular}{w{l}{2cm} cccc}
        \diagbox{\textbf{Training}}{\textbf{Testing}} 
        & LHC & Importance & \Block{}{Proposed\\Method} & Boundary  \\[0.8em] \cline{2-5}
        Unbalanced & 0.90 & 0.69 & 0.64 & 0.59 \\
        Balanced & 0.97 & 0.87 & 0.88 & 0.85 \\
        \bottomrule
    \end{NiceTabular}
\end{table}


\section{Summary of Findings and Discussion} \label{sec:discuss}

Generally, the proposed methods effectively obtains many samples within the HIC region, as depicted in \cref{fig:dataset_statistics}. Despite this high concentration of OPs centered around the security boundary, \cref{tab:dt_39} and \cref{tab:dt_162} demonstrate that DTs trained on this dataset exhibit good generalization to test sets generated by other methods, accurately classifying many points far from the security boundary. Moreover, as shown in \cref{fig:39_missclass} and \cref{fig:162_missclass}, misclassification frequently occurs near the security boundary, even by a DT trained with a high density of sampled points in the HIC region. These findings underscore the critical role of the security boundary in data-driven DSA methods and highlight its importance when designing synthetic datasets for such analyses.

Furthermore, \cref{fig:39_5_9} and \cref{fig:39_5_8} reveal that the security boundary is not always sharply defined. This lack of clarity may cause datasets generated by the proposed method and the importance sampling method to exhibit biases toward specific regions, even when the security boundary is broadly distributed. Such bias could explain why the LHC DT outperformed the importance DT on the proposed method’s test set, as shown in \cref{tab:dt_39}. Therefore, it is important to consider the security boundary in data-driven DSA, but to accurately describe the whole input space, a comprehensive dataset also needs OPs lying further away from the security boundary.

Additionally, the importance sampling method employed a multivariate normal distribution to sample data. While effective to some extent, this distribution may be insufficient for accurately capturing the complex structure of the security boundary. These results highlight the importance of selecting appropriate probability distributions or copulas \cite{konstantelos2018using} to better model and distinguish the marginal distributions that characterize the security boundary.

Finally, the proposed method effectively sampled many points in the HIC region and across the security boundary. However, its performance is influenced by several hyperparameters, including $\kappa_{max}$, $\kappa_{HIC}$, and $\epsilon$, as well as the location of the initialization points for the DW algorithm. These initialization points are crucial for accurately capturing the security boundary and generating a balanced dataset. Their location is guided by an effective mapping of the feasible region, facilitated by the systematic framework involving optimization-based bound tightening and the separating hyperplanes algorithm. Proper hyperparameter tuning and a sufficient number of initialization points are essential for the success of the proposed method in capturing the security boundary. Furthermore, when an unbalanced dataset is generated, a resampling step to balance the dataset can substantially improve the performance of a DT trained on this dataset.

\section{Conclusion} \label{sec:Conclusion}

Data-driven methods for dynamic security assessment (DSA) are gaining traction due to their significant computational speed advantages over conventional time-domain simulations. However, the performance of these tools heavily depends on the quality of the datasets used. Since historical records often lack sufficient coverage of the operating space, synthetic data generation methods become essential.

This paper proposes a novel method to generate synthetic datasets with a high concentration of operating points (OPs) around the security boundary. We demonstrate that datasets with a substantial share of OPs near the security boundary significantly enhance the performance of data-driven DSA tools. Our approach combines optimization-based bound tightening and infeasibility certificates derived from separating hyperplanes \cite{venzke2021efficient} to reduce the initial search space. We then employ directed walks \cite{thams2019efficient} to sample numerous OPs close to the security boundary.

We compare our proposed method against two benchmarks: a naive sampling benchmark and an importance sampling benchmark. Case studies on the PGLib-OPF 39- and 162-bus systems show that our method effectively samples a large number of points around the security boundary, capturing it more comprehensively than the benchmarks. Furthermore, decision trees (DTs) trained on the datasets generated by our method achieve a higher accuracy than those trained on benchmark-generated datasets. A closer examination of misclassified OPs reveals that misclassfication often occurs near the security boundary, emphasizing the need to accurately represent this region. Moreover, a comparison between DTs trained on balanced and unbalanced datasets highlights the importance of maintaining dataset balance while incorporating a substantial proportion of boundary samples, both of which are crucial for enhancing model performance.

Future work will focus on developing a unified sampling approach that combines the proposed method's ability to accurately describe the security boundary with copula-based importance sampling to efficiently generate large datasets. Additionally, we plan to evaluate the generated datasets for transient stability and extend the proposed method to other types of stability assessments, such as voltage stability.

\begingroup

\endgroup


\begin{thebibliography}{10}

\bibitem{panciatici2012operating}
P.~Panciatici, G.~Bareux, and L.~Wehenkel, ``Operating in the fog: Security management under uncertainty,'' {\em IEEE Power and Energy Magazine}, vol.~10, no.~5, pp.~40--49, 2012.

\bibitem{konstantelos2016implementation}
I.~Konstantelos, G.~Jamgotchian, S.~H. Tindemans, P.~Duchesne, S.~Cole, C.~Merckx, G.~Strbac, and P.~Panciatici, ``Implementation of a massively parallel dynamic security assessment platform for large-scale grids,'' {\em IEEE Transactions on Smart Grid}, vol.~8, no.~3, pp.~1417--1426, 2016.

\bibitem{zafar2018prosumer}
R.~Zafar, A.~Mahmood, S.~Razzaq, W.~Ali, U.~Naeem, and K.~Shehzad, ``Prosumer based energy management and sharing in smart grid,'' {\em Renewable and Sustainable Energy Reviews}, vol.~82, pp.~1675--1684, 2018.

\bibitem{duchesne2020recent}
L.~Duchesne, E.~Karangelos, and L.~Wehenkel, ``Recent developments in machine learning for energy systems reliability management,'' {\em Proceedings of the IEEE}, vol.~108, no.~9, pp.~1656--1676, 2020.

\bibitem{marot2022learning}
A.~Marot, B.~Donnot, K.~Chaouache, A.~Kelly, Q.~Huang, R.-R. Hossain, and J.~L. Cremer, ``Learning to run a power network with trust,'' {\em Electric Power Systems Research}, vol.~212, p.~108487, 2022.

\bibitem{sauer2007dynamic}
P.~W. Sauer, K.~L. Tomsovic, and V.~Vittal, ``Dynamic security assessment,'' {\em Power system stability and control}, vol.~5, pp.~421--430, 2007.

\bibitem{krishnan2011efficient}
V.~Krishnan, J.~D. McCalley, S.~Henry, and S.~Issad, ``Efficient database generation for decision tree based power system security assessment,'' {\em IEEE Transactions on Power Systems}, vol.~26, no.~4, pp.~2319--2327, 2011.

\bibitem{bugaje2023generating}
A.-A.~B. Bugaje, J.~L. Cremer, and G.~Strbac, ``Generating quality datasets for real-time security assessment: Balancing historically relevant and rare feasible operating conditions,'' {\em International Journal of Electrical Power \& Energy Systems}, vol.~154, p.~109427, 2023.

\bibitem{sevilla2022state}
F.~R.~S. Sevilla, Y.~Liu, E.~Barocio, P.~Korba, M.~Andrade, F.~Bellizio, J.~Bos, B.~Chaudhuri, H.~Chavez, J.~Cremer, {\em et~al.}, ``State-of-the-art of data collection, analytics, and future needs of transmission utilities worldwide to account for the continuous growth of sensing data,'' {\em International journal of electrical power \& energy systems}, vol.~137, p.~107772, 2022.

\bibitem{konstantelos2018using}
I.~Konstantelos, M.~Sun, S.~H. Tindemans, S.~Issad, P.~Panciatici, and G.~Strbac, ``Using vine copulas to generate representative system states for machine learning,'' {\em IEEE Trans. Power Syst.}, vol.~34, no.~1, pp.~225--235, 2018.

\bibitem{sun2016evaluating}
M.~Sun, I.~Konstantelos, S.~Tindemans, and G.~Strbac, ``Evaluating composite approaches to modelling high-dimensional stochastic variables in power systems,'' in {\em 2016 Power Systems Computation Conference (PSCC)}, pp.~1--8, IEEE, 2016.

\bibitem{liu2013systematic}
C.~Liu, K.~Sun, Z.~H. Rather, Z.~Chen, C.~L. Bak, P.~Th{\o}gersen, and P.~Lund, ``A systematic approach for dynamic security assessment and the corresponding preventive control scheme based on decision trees,'' {\em IEEE Trans. Power Syst.}, vol.~29, no.~2, pp.~717--730, 2013.

\bibitem{liu2013importance}
C.~Liu, Z.~H. Rather, Z.~Chen, C.~L. Bak, and P.~Th{\o}gersen, ``Importance sampling based decision trees for security assessment and the corresponding preventive control schemes: The danish case study,'' in {\em 2013 IEEE Grenoble Conference}, pp.~1--6, IEEE, 2013.

\bibitem{thams2019efficient}
F.~Thams, A.~Venzke, R.~Eriksson, and S.~Chatzivasileiadis, ``Efficient database generation for data-driven security assessment of power systems,'' {\em IEEE Trans. Power Syst.}, vol.~35, no.~1, pp.~30--41, 2019.

\bibitem{bugaje2023split}
A.-A.~B. Bugaje, J.~L. Cremer, and G.~Strbac, ``Split-based sequential sampling for realtime security assessment,'' {\em International Journal of Electrical Power \& Energy Systems}, vol.~146, p.~108790, 2023.

\bibitem{joswig2022opf}
T.~Joswig-Jones, K.~Baker, and A.~S. Zamzam, ``Opf-learn: An open-source framework for creating representative ac optimal power flow datasets,'' in {\em 2022 IEEE Power \& Energy Society Innovative Smart Grid Technologies Conference (ISGT)}, IEEE, 2022.

\bibitem{venzke2021efficient}
A.~Venzke, D.~K. Molzahn, and S.~Chatzivasileiadis, ``Efficient creation of datasets for data-driven power system applications,'' {\em Electric Power Systems Research}, vol.~190, p.~106614, 2021.

\bibitem{charles2024finalcode}
L.~Charles, ``Final code thesis lola charles.'' \url{https://github.com/lolachls/Final_Code_Thesis_Lola_Charles}, 2024.
\newblock Accessed: 2024-12-10.

\bibitem{boundtight}
C.~Coffrin, H.~L. Hijazi, and P.~Van~Hentenryck, ``Strengthening convex relaxations with bound tightening for power network optimization,'' in {\em Principles and Practice of Constraint Programming}, pp.~39--57, Springer International Publishing, 2015.

\bibitem{giraud_dsa_learn}
B.~Giraud {\em et~al.}, ``Dsa-learn.'' \url{https://github.com/bastiengiraud/DSA-learn}, 2024.
\newblock Accessed: 2024-11-30.

\bibitem{cigre2007review}
C.~W.~G. C4.601, ``Review of on-line dynamic security assessment tools and techniques,'' vol.~4, p.~601, 2007.

\bibitem{coffrin2015qc}
C.~Coffrin, H.~L. Hijazi, and P.~Van~Hentenryck, ``The qc relaxation: A theoretical and computational study on optimal power flow,'' {\em IEEE Transactions on Power Systems}, vol.~31, no.~4, pp.~3008--3018, 2015.

\bibitem{molzahn2019survey}
D.~K. Molzahn, I.~A. Hiskens, {\em et~al.}, ``A survey of relaxations and approximations of the power flow equations,'' {\em Foundations and Trends{\textregistered} in Electric Energy Systems}, vol.~4, no.~1-2, pp.~1--221, 2019.

\bibitem{kroese2013handbook}
D.~P. Kroese, T.~Taimre, and Z.~I. Botev, {\em Handbook of monte carlo methods}.
\newblock John Wiley \& Sons, 2013.

\bibitem{babaeinejadsarookolaee2019power}
S.~Babaeinejadsarookolaee, A.~Birchfield, R.~D. Christie, C.~Coffrin, C.~DeMarco, R.~Diao, M.~Ferris, S.~Fliscounakis, S.~Greene, R.~Huang, {\em et~al.}, ``The power grid library for benchmarking ac optimal power flow algorithms,'' {\em arXiv preprint arXiv:1908.02788}, 2019.

\bibitem{lara2021powersystems}
J.~D. Lara, C.~Barrows, D.~Thom, D.~Krishnamurthy, and D.~Callaway, ``Powersystems. jl—a power system data management package for large scale modeling,'' {\em SoftwareX}, vol.~15, p.~100747, 2021.

\bibitem{DTU_DCC_resource}
D.~C. Center {\em et~al.}, ``{DTU Computing Center resources},'' 2024.

\bibitem{coffrin2018powermodels}
C.~Coffrin, R.~Bent, K.~Sundar, Y.~Ng, and M.~Lubin, ``Powermodels. jl: An open-source framework for exploring power flow formulations,'' in {\em 2018 Power Systems Computation Conference (PSCC)}, IEEE, 2018.

\bibitem{dunning2017jump}
I.~Dunning, J.~Huchette, and M.~Lubin, ``Jump: A modeling language for mathematical optimization,'' {\em SIAM review}, vol.~59, no.~2, pp.~295--320, 2017.

\bibitem{wachter2006implementation}
A.~W{\"a}chter and L.~T. Biegler, ``On the implementation of an interior-point filter line-search algorithm for large-scale nonlinear programming,'' {\em Mathematical programming}, vol.~106, pp.~25--57, 2006.

\bibitem{lara2023powersimulationsdynamics}
J.~D. Lara, R.~Henriquez-Auba, M.~Bossart, D.~S. Callaway, and C.~Barrows, ``Powersimulationsdynamics. jl--an open source modeling package for modern power systems with inverter-based resources,'' {\em arXiv preprint arXiv:2308.02921}, 2023.

\bibitem{chalkis2020volesti}
A.~Chalkis and V.~Fisikopoulos, ``volesti: Volume approximation and sampling for convex polytopes in r,'' {\em arXiv preprint arXiv:2007.01578}, 2020.

\bibitem{breiman1984classification}
L.~Breiman {\em et~al.}, ``Classification and regression trees,'' {\em Monterey, CA: Wadsworth and Brooks/Cole}, 1984.

\bibitem{scikit-learn}
F.~Pedregosa, G.~Varoquaux, A.~Gramfort, V.~Michel, B.~Thirion, O.~Grisel, M.~Blondel, P.~Prettenhofer, R.~Weiss, V.~Dubourg, {\em et~al.}, ``Scikit-learn: Machine learning in {P}ython,'' {\em Journal of Machine Learning Research}, vol.~12, pp.~2825--2830, 2011.

\end{thebibliography}


\small
\end{document}